\documentclass[%
 reprint,
superscriptaddress,
 amsmath,amssymb,
 aps,
 prl,
]{revtex4-2}

\usepackage{graphicx}% Include figure files
\usepackage{dcolumn}% Align table columns on decimal point
\usepackage{bm}% bold math

\usepackage[utf8]{inputenc}
\usepackage{amsmath}
\usepackage{float}
\usepackage{setspace}
\usepackage{url}
\usepackage{enumitem}
\usepackage{siunitx}
\usepackage[T1]{fontenc}
\usepackage[bookmarks,colorlinks,citecolor=blue,linkcolor=blue]{hyperref}
\graphicspath{{Figures/}}

\begin{document}

\preprint{APS/123-QED}

\title{Alignment and anisotropy of stresses in disordered granular media}

\author{Aashish K. Gupta}
\email{aashish.gupta@ed.ac.uk}
\affiliation{School of Engineering, University of Edinburgh, Edinburgh EH9 3FG, United Kingdom}

\author{Christopher Ness}
\email{chris.ness@ed.ac.uk}
\affiliation{School of Engineering, University of Edinburgh, Edinburgh EH9 3FG, United Kingdom}

\author{Sina Haeri}
\email{s.haeri@hrwallingford.com}
\affiliation{School of Engineering, University of Edinburgh, Edinburgh EH9 3FG, United Kingdom}
\affiliation{HR Wallingford, Howbery Park, Wallingford, Oxfordshire, OX10 8BA, United Kingdom}

\date{\today}% It is always \today, today,

\begin{abstract}
Characterizing the degeneracy of local stress states is a central challenge in obtaining the complete statistical mechanics of disordered media. 
Here, we introduce a minimal force-balance model for isolated granular clusters to probe the structure of the stress space through principal stress orientation and stress anisotropy. We further show that when complemented by physically motivated pairwise constraints, the model produces predictions for the stress alignment in packings of repulsive hard spheres. 
We compare these predictions against simulation data for grains in hopper and simple shear flows, finding quantitative agreement.
This demonstrates the promise of modeling bulk athermal disordered systems through the combinatorics of few primitive geometric motifs.
\end{abstract}

\maketitle

Granular packings pose unique challenges for statistical mechanics due to their athermal and dissipative nature.
Traditional energy-based approaches fail, necessitating new statistical frameworks.
\citet{edwards1989theory} introduced compactivity $X = \partial V/\partial S$ as a temperature analog for grains, where $V$ is the system volume and $S$ is the entropy associated with the number of possible configurations.
This quantifies the ability of the system to accommodate volumetric changes~\citep{lechenault2006free,mcnamara2009measurement},
enabling prediction of packing densities and volume fluctuations in dense assemblies~\citep{aste2008emergence,song2008phase,hihinashvili2012statistical}.
However, compactivity focuses solely on volume, neglecting the force networks crucial for mechanical stability~\citep{henkes2009statistical,bi2015statistical}.
To address this,
\citet{blumenfeld2009granular} introduced angoricity $\hat{\alpha} = \partial S/ \partial\hat{\Sigma}$, a tensorial analog of inverse temperature with $\hat{\Sigma}$ the global force-moment tensor.
The probability of observing a granular cluster of local force-moment $\hat{\sigma}$ is then
$P(\hat{\sigma}| \,\hat{\alpha})= (\Omega(\hat{\sigma})/Z(\hat{\alpha}))\,\exp(-\hat{\alpha}:\hat{\sigma})$,
where
$\Omega(\hat{\sigma})$ is the degeneracy in the stress state (innate frequency of occurrence in isolation), and $Z(\hat{\alpha},\kappa)$ is the partition function.
The comprehensive form of $\Omega$ remains elusive,
however,
and conventional formalisms~\cite{bililign2019protocol,rey2021procedure,baranau2024estimating}
characterize $\hat{\sigma}$ using two scalar invariants of the stress, overlooking a crucial implicit physical invariant, namely the principal stress direction.
These approaches thus predict the distribution and fluctuations of the macroscopic stress, 
but fail to capture the spatial organization of the stress fields.
We argue that incorporating the relative orientation of principal stresses of contacting particles is essential in recovering this missing spatial structure.

Here, we advance the understanding of stress transmission in granular media in three directions.
First, we study an isolated cluster of identical grains and identify all theoretically admissible configurations satisfying equilibrium conditions.
For each configuration, we compute the stress anisotropy and the orientation of the principal stress directions,
making progress toward the question of the degeneracy of stress states.
Second, we quantify the misalignment of the stress transmission directions for contacting particles in a cluster.
This is key to reimagining the current statistical mechanics framework for force chain predictions.
Finally, we address hopper and simple shear flow by particle simulation,
as exemplars of real granular dynamics.
We compare the stress distributions thus obtained to those of the isolated cluster,
demonstrating that the primitive case contains sufficient physics to predict bulk behavior.
As well as advancing the statistical mechanics of granular matter, our work guides the development of force chain screening algorithms, a mainstay of industrial granular technology research.
\begin{figure}
    \centering
    \includegraphics[width=\linewidth]{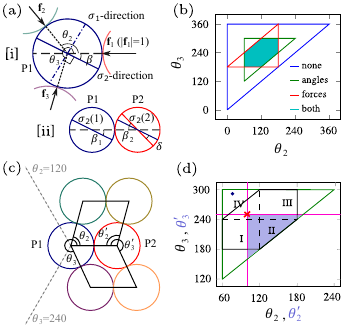}
    \caption{
    Isolated Frictionless Cluster~(IFC) model for stress transmission in 2D.
    (a)[i] Disk $P1$ acted on by forces ${\bm f}_i$ at angles $\theta_i$ measured counterclockwise from the outward normal at the first contact;
    the blue line making acute angle $\beta$ with the outward normal
    and the dash-dotted perpendicular line show the minor $\sigma_2$ and major $\sigma_1$ principal stress directions;
    [ii] misalignment angle $\delta$ between $\sigma_2$ of $P1$ and P2;
    (b) subregions of the $\theta_2-\theta_3$ space satisfying:
    equilibrium conditions (blue),
    minimum angular separations of $60$ (green), 
    inward directionality of forces (red),
    constrained angles and forces (turquoise);
    (c)
    example configuration with $P1$ and $P2$ neighbors located at $(\theta_2,\theta_3)$ and $(\theta_2',\theta_3')$ respectively.
    Dashed lines demarcate the constrained Regions $I$-$IV$ described in the text;
    (d) angular configuration space showing $(\theta_2,\theta_3)$ from (c) (blue dot) lies in R$IV$, demanding that $(\theta_2',\theta_3')$ (red cross) lies in, or on the boundary of, the shaded region.
    }
    \label{fig_1}
\end{figure}

\textit{Model.}---We consider first a frictionless disk $P1$
of volume $v$ in static equilibrium experiencing contact forces from three identical neighbors, Fig.~\ref{fig_1}(a)[i].
The first applies a unit inward force $\bm{f}_1$.
A second force $\bm{f}_2$
acts at a rotation of $\theta_2$ from the outward normal at $\bm{f}_1$,
and $\bm{f}_3$ at $\theta_3$.
Forces $\bm{f}_2$ and $\bm{f}_3$ are thus uniquely determined for $P1$ in equilibrium.
By incrementally varying $\theta_2$ and $\theta_3$,
we construct for each realization the force-moment tensor
$\hat\Sigma =\sum_{i=1}^{3}\bm{f}_i \otimes \bm{r}_i$,
with $\bm{r}_i$ the contact points.
We then define $\sigma:=\hat\Sigma/v$,
which is related to the local Cauchy stress as $\sigma=\frac{1}{v} \int_v \sigma_c \,dv$
and is thus symmetric and has real eigenvalues and eigenvectors.
The direction of one eigenvector $\gamma$ is
$\tan 2\gamma=2\sigma_{12}/(\sigma_{11}-\sigma_{22})=(f_{2} \sin 2\theta_2 + f_{3} \sin 2\theta_3)/ (1+ f_{2} \cos 2\theta_2 + f_{3} \cos 2\theta_3)$,
with the other orthogonal.
The orientation of the minor (most compressive) principal stress $\sigma_2$ follows the eigenvector of the lower eigenvalue of $\sigma$.
The stress inclination $\beta$ for $P1$ is the acute angle between $\sigma_2$ and the outward normal at $\bm{f}_1$, counterclockwise being positive.
We define the anisotropy $A=(\sigma_1-\sigma_2)/(\sigma_1+\sigma_2)$
as the ratio of the maximum shear to hydrostatic stress.
The values of $\beta$ and $A$ are invariant to changes in $\bm{f}_1$ by a positive prefactor $c$;
for negative $c$, the minor and major principal axes are swapped,
transforming $\beta$ and $A$ accordingly. 

Figure~\ref{fig_1}(b) illustrates subregions of the $\theta_2-\theta_3$ parameter space accessible under physically motivated constraints on $\theta_{2,3}$ and $\bm{f}_{2,3}$.
The region satisfying force equilibrium is bounded by the blue triangle with hypotenuse
$\theta_3=\theta_2$
($\theta_3\ge \theta_2$ by construction).
The green triangle is defined by $\theta_3\geq\theta_2+60$, $\theta_2\geq60$ and $\theta_3\leq300$,
precluding particle overlaps.
Red lines bound the region where only inward forces are permitted,
representing noncohesive particles, with $0\leq\theta_2\leq180$,  $\theta_3\geq180$ and $\theta_3\leq\theta_2+180$.
The turquoise intersection satisfies all constraints.

\begin{figure}[b]
    \centering
    \includegraphics[width=\linewidth]{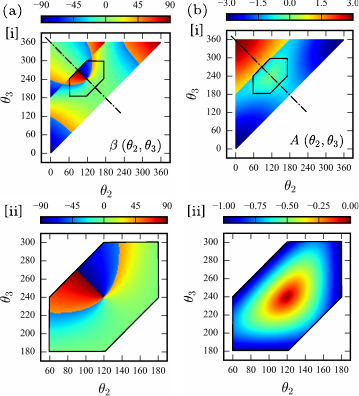}
    \caption{
    Variation over $\theta_2-\theta_3$ parameter space of
    (a)[i] the minor principal stress angle $\beta$,
    and
    (b)[ii] the anisotropy $A$.
    Dash-dotted line shows $\theta_2 + \theta_3=360$.
    The hexagonal regions (magnified in (a)[ii] and (b)[ii]) satisfy constraints on the minimum angular separation between neighbors and inward directionality of the forces.
    }
    \label{fig_2}
\end{figure}

\textit{Results.}---We first describe the features of $\beta$ and $A$ for positive $\bm{f}_1$,
Figs.~\ref{fig_2}(a)[i] and (b)[i].
Notably,
$\beta$ is antisymmetric whereas $A$ is symmetric about $\theta_2+\theta_3=360$,
since
reflecting $P1$ about the horizontal transforms $(\theta_2,\theta_3)$ to $(360-\theta_3,360-\theta_2)$,
leaving the stress state unchanged.
A striking consequence of this is a discontinuity in $\beta$, jumping from $-90$ to $90$ across the symmetry line.
Though abrupt, these values actually correspond to the same vertical orientation,
unlike other $\pm \beta$ pairs.
We observe other sets of curves across which $\beta$ jumps from $-45$ to $45$,
satisfying $1+f_2\cos2\theta_2 + f_3\cos2\theta_3=0$.
Taking stresses positive in tension,
$A<0$ indicates particle compression.
A particularly sensitive configuration arises when $\bm{f}_3$ diametrically opposes $\bm{f}_2$ ($\theta_3 = \theta_2 + 180$).
Here a perturbation $\theta_3 + \theta_h$ transforms $\bm{f}_3$ from compressive to tensile.
This results in a sharp transition from $A = -1$ when $|\sigma_2| \gg |\sigma_1|$ (compression), to $A = 1$ when $|\sigma_1| \gg |\sigma_2|$ (tension),
across the aforementioned line.
In the limiting cases corresponding to the triangle vertices in Fig.~\ref{fig_2}(a)[i] and (b)[i],
$A$ approaches $\pm3$, corresponding to $\sigma_1=-2\sigma_2$ and $\sigma_2=-2\sigma_1$ respectively.

We focus next on the black hexagon in Fig.~\ref{fig_2}.
Along its lower ($\theta_3=180$) and right ($\theta_2=180$) edges,
the entire force countering $\bm{f}_1$ originates from a particle in line with it.
The
minor stress then aligns so that $\beta=0$,
with
$\sigma_1=0$ and consequently $A=-1$.
In the interior of the hexagon,
we find the steepest gradient in $A$ is along $\theta_2+\theta_3=360$.
The point ($\theta_2$,$\theta_3$)=(120,240) on the line is special in the sense that it is maximally symmetric, making it the global minimum of $|A|$ with $\sigma_1=\sigma_2$ and $A=0$.
Small perturbations about this point drastically alter the principal stress direction.

\begin{figure}
    \centering
    \includegraphics[width=\linewidth]{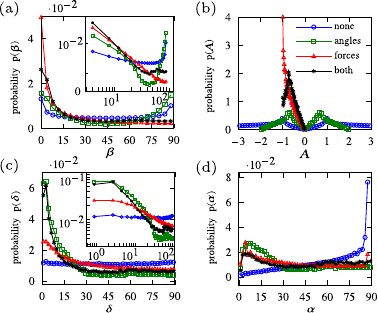}
    \caption{Probability density functions of
    (a) $\beta$,
    (b) $A$,
    (c) $\delta$ and
    (d) $\alpha$
    obtained from the IFC model
    for the different physically motivated constraints (none, angles, forces, both) on the configuration and contact forces.
    Insets in (a) and (c) show log-log plots of the same quantities as their main panels.
    }
    \label{fig_3}
\end{figure}

After evaluating $\beta$ over the $\theta_2$–$\theta_3$ space,
we compute its probability density $p(\beta)|_{c=1}$,
Fig.~\ref{fig_3}(a).
Since $\beta$ is unchanged for positive $c$,
$p(\beta)|_{\forall c>0}$ is identical to $p(\beta)|_{c=1}$.
For $c<0$
the principal axes are swapped,
implying $\beta \mapsto \beta - 90$ for $\beta > 0$, and $\beta \mapsto \beta + 90$ for $\beta < 0$.
Thus $p(\beta)|_{\forall c<0}$ can be obtained from $p(\beta)|_{\forall c>0}$,
inheriting the same symmetry about $\beta = 0$.
The full distribution $p(\beta)|_{\forall c}$
averaged across arbitrary $c$
is therefore symmetric,
so Fig.~\ref{fig_3}(a) reports $\beta \geq 0$,
and there is an emergent symmetry about $\beta = 45$ when tensile forces are allowed (blue and green lines).
The inset of Fig.~\ref{fig_3}(a) shows the same data on $\log$ axes, revealing a power-law decay under force constraints,
whereas
unconstrained cases additionally exhibit a rise toward the extrema of $\beta$.

Similar considerations apply to $p(A)$, Fig.~\ref{fig_3}(b).
Unlike $\beta$, which transforms non-trivially under sign changes in $c$,
$A$ obeys $A|_{c=c'}=-A|_{c=-c'}$ when the principal stresses are interchanged,
since $\sigma_2|_{c=-c'}=-\sigma_1|_{c=c'}$ and $\sigma_1|_{c=-c'}=-\sigma_2|_{c=c'}$.
Consequently, $p(A)$ is symmetric about $A = 0$ only when negative $c$ are included
(blue and green lines),
otherwise the asymmetry persists.
Under all constraints, $p(A)$ is nonmonotonic between $A=-1$ and $0$.
Under angle constraints, $A$ reaches $\pm2$ at the triangle vertices in Fig.~\ref{fig_1}(b),
whereas unconstrained $A$ reaches $\pm3$,
the extremes arising when angular separations vanish.
Notably, these cases exhibit positive anisotropy since tensile forces are permitted on the periphery as valid solutions.
These symmetries in the $P1$ stress state
can guide the form of $\Omega(\hat{\sigma})$ in the statistical mechanics formalism.

\begin{figure*}
    \centering
    \includegraphics[width=\linewidth]{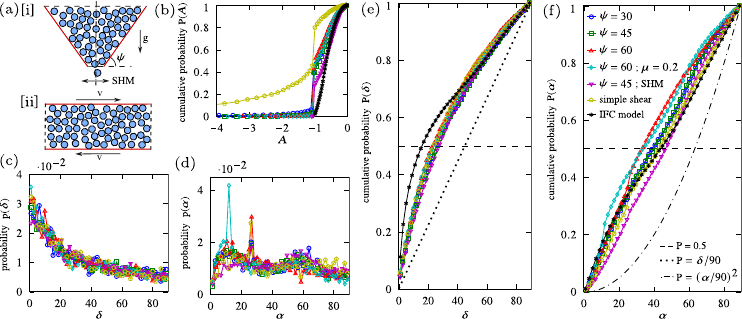}
    \caption{
    Comparison of IFC model results with particle simulations.
    Shown are
    (a)[i] gravity-driven hopper flow with outlet angle $\psi \in \{30, 45, 60\}$,
    and
    (a)[ii] simple shear in the absence of gravity,
    with dashed lines representing periodic boundaries.
    Shown in (b) is the cumulative distribution function $P(A)$ for the simulated cases and the IFC model.
    (c) and (d) show $p(\delta)$ and $p(\alpha)$,
    with the corresponding $P(\delta)$ and $P(\alpha)$
    shown respectively in (e) and (f).
    The legend for (b)-(f) is given in (f).
    The dashed horizontal lines in (e) and (f) separate the top and bottom half of the distribution to identify the critical angles,
    while the dotted and dash-dotted curves are the hypothetical $P(\delta)$ and $P(\alpha)$ assuming $\beta$ were uniformly distributed over $[-90,90]$.}
    \label{fig_4}
\end{figure*}

Our analysis naturally extends to disk pair $P1$-$P2$, Fig.~\ref{fig_1}(a)[ii],
unveiling the relative orientation of their principal stresses.
Let $\beta$ take values $\beta_1$ for $P1$ and $\beta_2$ for $P2$,
with the latter experiencing $-\bm{f}_1$ and two other forces at $\theta_2^\prime$, $\theta_3^\prime$ (measured clockwise).
Given $\beta_1$ and $\beta_2$, we calculate the acute misalignment $\delta$ between the minor principal stresses as
$\delta = |\beta_1 + \beta_2|$,
reverting to $\delta = 180 - |\beta_1 + \beta_2|$
when
$|\beta_1 + \beta_2| > 90$.
As per the forward-backward criteria put forth in~\cite{peters2005characterization},
to be classified as integral to a force chain,
a branch vector must lie within a given angle of $\sigma_2$ of each particle.
The angle $\alpha:=\max \,(|\beta_1|, |\beta_2|)$ thus screens for neighbors that are potentially important for mechanical stability.

The allowed neighbor configurations for $P2$ depend on $P1$ neighbor placements.
In the example shown in 
Fig.~\ref{fig_1}(c),
the $P1$ neighbors have
$\theta_2 < 120$ and $\theta_3>240$,
so under angle constraints, the neighbors of $P2$ must obey
$\theta_2'\geq180-\theta_2$ and
$\theta'_3 \leq (360 - \theta_3) + 180$,
indicated by the highlighted region of Fig.~\ref{fig_1}(d).
Shown in Fig.~\ref{fig_1}(c) is the extreme case where
$\theta_2'=180-\theta_2$ and
$\theta'_3 = (360 - \theta_3) + 180$.
There are distinct constraint classes for $(\theta_2',\theta_3')$, defined according to the following quadrants R$I$-$IV$ of the 
$\theta_2-\theta_3$ parameter space.
\textit{Region I}: \( \theta_2 \leq 120 \), \( \theta_3 \leq 240 \) \(\Rightarrow\) \( \theta'_2 \geq 180 - \theta_2 \);
\textit{RII}: \( \theta_2 > 120 \), \( \theta_3 \leq 240 \) \(\Rightarrow\) no additional constraints;
\textit{RIII}: \( \theta_2 > 120 \), \( \theta_3 > 240 \) \(\Rightarrow\) \( \theta'_3 \leq (360 - \theta_3) + 180 \);
\textit{RIV}: \( \theta_2 \leq 120 \), \( \theta_3 > 240 \) \(\Rightarrow\) both \( \theta'_2 \geq 180 - \theta_2 \) and \( \theta'_3 \leq (360 - \theta_3) + 180 \).

 Without constraints, all $\delta$ are roughly equiprobable,
 Fig.~\ref{fig_3}(c).
 Force constraints skew $p(\delta)$ toward small $\delta$,
 as do angle constraints.
 When both are applied, the distribution closely matches that with angle constraints alone.
 The power-law decay of $p(\delta)$ (Fig.~\ref{fig_3}(c) inset)
 perhaps underpins phenomena such as power-law rheology across diverse geometries~\citep{kim2020power},
power-law energy spectra in dense flows~\citep{oyama2019avalanche}, and depth-dependent power-law decay of wave characteristics following surface impacts~\citep{hong1999power}.
 Similar scale-invariance appears in seismic phenomena,
 including the frequency-magnitude distribution of earthquakes and temporal decay of aftershock rates~\citep{narteau2009common,meng2019power}.
The unconstrained $p(\alpha)$ increases linearly
 except near $\alpha=90$, Fig.~\ref{fig_3}(d), where the surge is attributed to contributions from  $c<0$ cases.
 Under angle and force constraints, $p(\alpha)$ peaks at $\approx5$ before plateauing for $\alpha\gtrsim30$.
The statistics obtained thus far motivate an extension of the angoricity framework,
wherein an orientational misalignment angle supplements the scalar stress invariants as a measure of the structural properties of force-carrying contacts.

\textit{Simulation.}---To demonstrate the utility of the IFC model,
we compare the predicted
cumulative probability distributions $P(A)$, $P(\delta)$ and $P(\alpha)$ to those obtained by 2D particle simulation of granular processes, Fig.~\ref{fig_4}(a).
We use a damped Hookean contact model~\cite{cundall1979discrete,kloss2012models} to evolve trajectories during periodic gravity-driven hopper, and simple shear flow,
with stiffnesses
$k_n=10^6 mg/d_p$ and
$k_t=2k_n/7$,
damping
$\gamma_n=317 m \sqrt{g/d_p}$
and
$\gamma_t=\gamma_n/2$,
density $\rho$,
and friction coefficient $\mu=0.5$ unless stated otherwise.
 The collision time is $t_c=\pi (2k_n/m-\gamma_n^2/4m^2)^{-1/2}$,
 the timestep is $0.01t_c$
 and 
 we simulate $\sim5 \times 10^3$ particles (an equimolar mixture of diameters $0.95 d_p$, $d_p$, $1.05 d_p$).
 For hopper flow
 the particles are seeded above a geometry defined by
 wall inclination $\psi$, edge length $90d_p$, and opening width $30d_p$
 (preventing clogging).
 We also simulate a strongly agitated hopper under simple harmonic motion (SHM) with amplitude $8d_p$ and period $640 t_c$.
For simple shear the
nominal rate is $\dot{\gamma}/(k_n/\rho d_p^3)^{1/2}= 8.23 \times10^{-6}$
and the area fraction is $\phi =0.82$.
We extract principal stress data using a recursive-force-chain algorithm~\cite{ejtehadi2024force}, applying a threshold $\alpha_\mathrm{max}=90$ to include all contacting particles.

Figure~\ref{fig_4}(b) shows $P(A)$ for each case (legend in Fig.~\ref{fig_4}(f)).
The jump at $A=-1$ ($\sigma_1$=0) indicates
a single dominant stress transmission direction,
and its presence even in shear implies gravity is not solely responsible.
It likely relates to a dominant compressive axis of load transmission under dense conditions~\cite{lois2007spatial}, whereas the IFC model permits greater homogeneity in the distribution of stress.
For noncohesive frictionless particles both principal stresses are compressive,
so $\sigma_1/\sigma_2>0$ as for the IFC model.
A local pure shear would produce compressive and extensive stresses of equal magnitude,
giving $\sigma_1/\sigma_2=-1$ and $A\to-\infty$.
This doesn't occur in simulations because frictional forces can't exist without accompanying normal forces,
but the combined effect might nonetheless produce a tensile $\sigma_1$.
This occurs rarely, however, as demonstrated by the low $P(A)$ when $A<-1$,
except in simple shear where tangential and normal forces are comparable.
For the SHM case, $A>-1$ due to the dominance of compressive forces as a result of heavy oscillation.

The distribution $p(\delta)$ decays rapidly with $\delta$,
collapsing across all cases Fig.~\ref{fig_4}(c).
As shown in Fig.~\ref{fig_4}(d),
however, $\alpha$ is more sensitive to details,
with $p(\alpha)$ exhibiting
peaks at $\alpha \approx \tan^{-1}(\mu)$ due to the full mobilization of friction at sliding contacts.
The corresponding $P(\delta)$ and $P(\alpha)$ are shown in Figs.~\ref{fig_4}(e) and (f),
with IFC results superposed. 
Assuming \( \beta_1, \beta_2 \sim \mathcal{U}[-90, 90] \),
the cumulative distribution \( P(\delta) \) follows from \( P(|\beta_1 + \beta_2|) \). Convolution of \( p(\beta_1) \) and \( p(\beta_2) \) yields a triangular distribution on \([-180, 180]\), which folds onto \([0, 180]\) via the modulus. Values in \([90, 180]\) reflect about \( |\beta_1 + \beta_2| = 90 \) into \([0, 90] \), yielding a uniform distribution \( P(\delta = \delta') = \delta'/90 \).
Likewise, for \( \alpha = \max(|\beta_1|, |\beta_2|) \), the cumulative distribution is \( P(\alpha = \alpha') = P(|\beta_1| = \alpha') \cdot P(|\beta_2| = \alpha') = (\alpha'/90)^2 \).
The discrepancy between the uniform-$\beta$ distributions and those from the model and simulation demonstrates the non-triviality of the results,
whereas the broad agreement across our results highlights universality in $\delta$ and $\alpha$ that is captured by the primitive components of our model.

Notably, the IFC model captures the statistics even when gravity is present.
Inserting a body force $mg$ into Fig.~\ref{fig_1}(a)[i],
the probability density of a stress state of given $\beta$ and $A$ becomes
$p(\beta, A)=1/C_L \left[ \int_0^{kmg} p(A,\beta)|_c \, dc + \int_{kmg}^{C_L} p(A,\beta)|_c \, dc \right]$.
Here, the first integral is over forces similar in magnitude to $mg$
($k$ is an $\mathcal{O}(1)$ constant with units of inverse force)
and the second is over larger forces up to a maximum set by $C_L$.
When $C_L\gg kmg$,
the latter integrand is $c$-invariant and approximates the gravity-free distribution.
This scenario is supported by our simulation data,
for which contact forces
typically exceed $mg$ by an order of magnitude.
Furthermore, while the distribution of $A$ varies with experimental details,
the angles are sufficiently robust
that prescribing $\alpha_\mathrm{max}\approx 33-46$ (where $P(\alpha)=0.5$, Fig.~\ref{fig_4}(f)) one could successfully identify load-bearing particles~\citep{peters2005characterization,ejtehadi2024force}.

\textit{Conclusions.}---Overall, the agreement between IFC statistics and simulations of bulk systems underscores the abundance of redundant information in the latter, which is encouraging from a constitutive modeling standpoint.
The observed scale-invariant relationships in the macroscopic granular response likely originate from the power-law distribution of $\delta$, which we show to be universal.
The IFC model can be extended naturally to deformable tissues, cells, or foams by reducing the minimal angular separation according to the stiffness,
whereas an upper bound on tensile forces could be imposed in the equilibrium solution depending on cohesion. While the current model excludes larger contact numbers due to static indeterminacy~\citep{unger2005force,mcnamara2005indeterminacy,behringer2018physics},
future work may overcome this limitation and generalize the analysis to markedly polydisperse disordered systems.

\begin{acknowledgments}
A.K.G.~thanks Omid Ejtehadi, Kevin Hanley, Anushanth Karalasingam, and Ravi Gautam for helpful discussions, and the University of Edinburgh for supporting this work through a doctoral studentship.
C.N.~acknowledges support from the Royal Academy of Engineering under the Research Fellowship scheme and from the Leverhulme Trust under Research Project Grant RPG-2022-095.
\end{acknowledgments}

\bibliography{apssamp}% Produces the bibliography via BibTeX.

\end{document}